\documentclass[aps,twocolumn,pra,nopacs]{revtex4-1}

\usepackage{latexsym}
\usepackage{amsmath}
\usepackage{natbib}
\usepackage[pdftex]{graphicx}
\usepackage{epstopdf}
\usepackage{times}
\usepackage{txfonts}
\usepackage{xcolor}
\usepackage{textcomp}
\usepackage[colorlinks, bookmarks=false, citecolor=blue, linkcolor=red,urlcolor=blue]{hyperref}

\begin{document}

\title{Physical realization of the generalized fully
frustrated XY model in an array of SFS junctions}

\author{S.~E. Korshunov}
\affiliation{L.~D.~Landau Institute for Theoretical Physics RAS, 142432
Chernogolovka, Russia}

\date{\today}


\begin{abstract}
We show that a physical realization of the phase diagram proposed by Minnhagen {\em et al.} [Phys. Rev. B 
{\bf 78}, 184432 (2008)] for so-called generalized fully frustrated XY model on square lattice can be achieved in arrays of SFS (superconductor-ferromagnet-superconductor) junctions near the transition of the junctions to $\pi$-state. Moreover, the phase diagram with such a topology has to be reproduced {\em twice}, on both sides of the $0-\pi$ transition.

\end{abstract}

\maketitle
\section{Introduction}

A fully frustrated (FF) XY model can be defined by the Hamiltonian
\begin{subequations}                                         \label{FFXY}
\begin{equation}
H=\sum_{(jj')}U(\theta_{j'}-\theta_{j}-A_{jj'})\;,          \label{H}
\end{equation}
where 
the form of the interaction of variables $\theta_{j}$  defined on sites
$j$ of some regular two-dimensional lattice is described by an even periodic function $U(\phi)$ (with period $2\pi$) minimal at $\phi=0$. The summation in Eq. (\ref{H}) is performed over all
pairs $(jj')$ of nearest neighbors on the lattice. The non-fluctuating
(quenched) variables $A_{jj'}\equiv -A_{j'j}$ defined on lattice bonds
have to satisfy the constraint
\begin{equation}                                            \label{f}
\sum_{\mbox{\small\raisebox{-0.6mm}{$\Box\,$}}} A_{jj'}=\pm\pi
\,(\mbox{mod}\,2\pi)
\end{equation}
\end{subequations}
on all  lattice plaquettes. The notation
$\mbox{\raisebox{-0.4mm}{$\Box$}}\hspace*{0.7mm}$ below the sign of
summation implies the directed sum of variables $A_{jj'}\equiv -A_{j'j}$
over the perimeter of a plaquette in the counterclockwise
direction. The model is gauge-invariant, that is, its properties do not depend on the particular choice of gauge variables $A_{jj'}$ as long as constraint (\ref{f}) is satisfied.

The model defined by Eqs. (\ref{FFXY}) with interaction function
\begin{equation}                                              \label{U0}
    U(\phi)=U_0(\phi)=J(1-\cos\phi)\;
\end{equation}
can be used for the description of a planar magnet with odd number of antiferromagnetic bonds per plaquette 
\cite{Villain77} and of a magnetically frustrated Josephson junction array with a half-integer number of flux quanta per plaquette \cite{Teitel83B}. In the latter case, $\theta_j$ is the phase of the order parameter on the $j$th superconducting island and \makebox{$\phi_{jj'}\equiv\theta_{j'}-\theta_{j}-A_{jj'}$}
the gauge-invariant phase difference between the neighboring islands.
Note that here and below $U(\phi)$ is always assumed to be counted off from its value at $\phi=0$, where $U(\phi)$ is minimal.

In contrast to the case of the FF XY model on dice lattice \cite{Korshunov01}, the ground states of the FF XY model on square or triangular lattice can be constructed by minimizing the energy of each plaquette independently. For a square plaquette, the minimization of
\begin{equation}                                             \label{Eplaq4}
    E_{\rm plaq}=\sum_{\alpha=1}^{4} U(\phi_\alpha)
\end{equation}
under the constraint $\sum_{\alpha=1}^{4}\phi_\alpha=\pm\pi$ following from Eq. (\ref{f}) for $U(\phi)=U_0(\phi)$ gives
\begin{equation}                                        \label{phialpha}
 \phi_\alpha=\pm\pi/4\;,
\end{equation}
which immediately allows one to construct a ground state with a checkerboard structure which in addition to continuous $U(1)$ degeneracy (related to the simultaneous rotation of all variables $\theta_j$) has also a two-fold discrete degeneracy related to the choice of sign in Eq. (\ref{phialpha}) \cite{Villain77}.

In accordance with that, the model allows for the existence of two phase transitions related to the disordering of the continuous and discrete degrees of freedom (see Refs. \onlinecite{Hasenbusch05} and \onlinecite{Korshunov06r,*Korshunov06} for reviews). The former takes place at slightly lower temperature than the latter
due to the screening of the interaction of the conventional (integer) vortices by the kinks on domain walls behaving themselves as fractional vortices \cite{Korshunov02}.

Few years ago Minnhagen {\em et al.} \cite{Minnhagen07,Minnhagen08} suggested that a substantial change in the form of the interaction function in Eq. (\ref{H}) may lead to a change in the sequence and/or nature of phase transitions in the system. They chose to replace $U_0(\phi)$  by the interaction function (\ref{Ugen})
introduced earlier \cite{Domany84,vanHimbergen84PRL} for studying the transformation of the phase transition in the conventional
XY model from a continuous to a first-order one. The phase diagram of the FF XY model with such an interaction (named the generalized FF XY model) has been investigated in Refs. \onlinecite{Minnhagen07} and \onlinecite{Minnhagen08} by numerical methods. It has a rich structure with four different phases separated by transition lines of various types and incorporates four multicritical points.

The main aim of the present note is to point out that a physical realization of the phase diagram proposed by Minnhagen {\em et al.} for the generalized FF XY model on square lattice can be achieved in arrays of SFS (superconductor-ferromagnet-superconductor) junctions \cite{Ryazanov01,Golubov04,Frolov08} near the transition of the junctions to so-called $\pi$-state \cite{Chtch01r,*Chtch01,Barash02,Golubov02r,*Golubov02}. After analyzing in Sec. II the main reasons for the overall structure of the phase diagram of the generalized FF XY model, in Sec. III we demonstrate that in the most relevant range of parameters the function describing the energy of a SFS junction has almost the same form as the one used in Refs. \onlinecite{Minnhagen07} and \onlinecite{Minnhagen08}.
As a consequence, a magnetically frustrated array of SFS junctions with half-integer number of flux quanta per plaquette has to demonstrate a rich phase diagram with the structure analogous to that of the generalized FF XY model.

\section{Generalized FF XY model}

As has been already mentioned in the Introduction, in Refs. \onlinecite{Minnhagen07} and \onlinecite{Minnhagen08} Minnhagen {\em et al.} investigated the FF XY model defined by Eqs. (\ref{FFXY})
with interaction function,
\begin{equation}                                                  \label{Ugen}
    U_{\rm gen}(\phi) =\frac{2J}{p^2}
    \left\{1-\left[\cos^{2}\left(\frac{\phi}{2}\right)\right]^{p^2}\right\}\,.
\end{equation}
This interaction function had been introduced in Ref. \cite{Domany84} to study the influence of the form of the interaction (depending on the value of parameter $p$) on the type of the phase transition in the conventional (without frustration) XY model.
At $p=1$ function $U_{\rm gen}(\phi)$ defined by Eq. (\ref{Ugen}) coinsides with $U_0(\phi)$. The $p$-dependent numerical prefactor in the right-hand side of Eq. (\ref{Ugen}) had been introduced to make $U''(0)=J$ independent of $p$. This parameter determines the energies of smooth continuous fluctuations (spin waves and vortices) in the unfrustrated case. The FF XY model with interaction function (\ref{Ugen}) was named in Ref. \onlinecite{Minnhagen07} the generalized fully frustrated XY model.

\begin{figure}[b] (a)
\includegraphics[width=7cm]{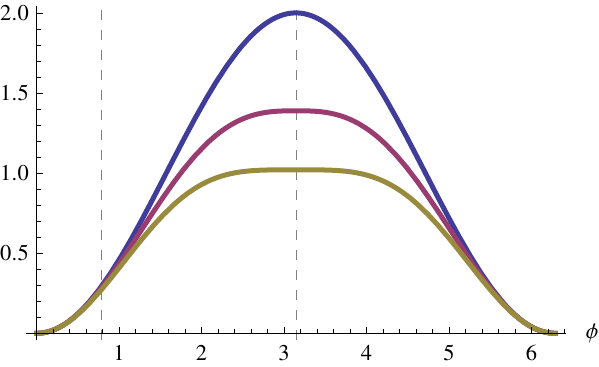}

(b)
\includegraphics[width=7cm]{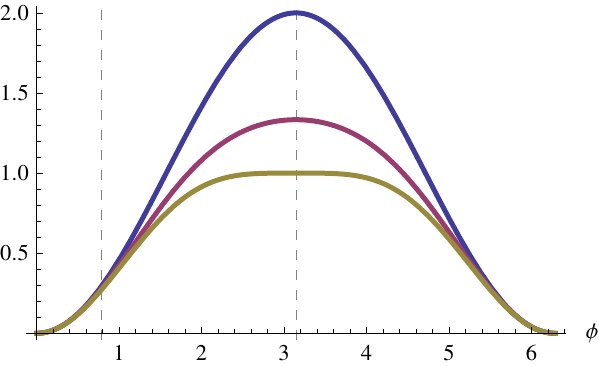}
\caption{(color online) (a) Comparison of the form of $U_{\rm gen}(\phi)$ for $\mbox{p=1}$ (upper curve), $p=1.2$ and $p=1.4$ (lower curve); (b) comparison of the form of $U_{\rm SFS}(\phi)$ for $t=0$ (upper curve), $t=0.125$ and $t=0.25$ (lower curve). Vertical dashed lines correspond to $\phi=\pi/4$ and $\phi=\pi$. In this and all other plots the energies of a bond (or of a plaquette) are given in units of coupling constant $J$.
}
\label{fig:comparison}
\end{figure}

The most important property of the generalized FF XY model on square lattice
(responsible for the overall structure of its phase diagram) is the change in the structure of its ground states \cite{Minnhagen07}, which takes place with the increase in $p$ at
\begin{equation}                                                   \label{pc}
    p_c=\sqrt{\frac{\ln(3/4)}{2\ln[\cos(\pi/8)]}}\approx 1.347880\;.
\end{equation}
At $p<p_c$ the ground states of the generalized FF XY model have the same structure as in the conventional one (corresponding to $p=1$). However, at $p>p_c$ the energy of a single plaquette is minimized when on one bond $\phi=\pi$, whereas on three others bonds $\phi=0$.

In order to explain the reasons for the modification of the ground state structure in the generalized FF XY model with the increase of $p$, Fig. \ref{fig:comparison}(a) compares the form of $U_{\rm gen}(\phi)$ for three different values of parameter $p$ (${p=1}$, $p=1.2$ and $p=1.4$). It is evident from this plot that the increase of $p$ suppresses the value of $U_{\rm gen}(\pi)$ much stronger than that of $U_{\rm gen}(\pi/4)$. This suggests that instead of uniformly distributing the gauge-invariant phase difference $\pm\pi$ between the four bonds belonging to a plaquette it can become more advantageous to concentrate it completely on one bond with three others having the minimal energy, that is $\phi=0$.

\begin{figure}[t] (a) 
\includegraphics[width=7cm]{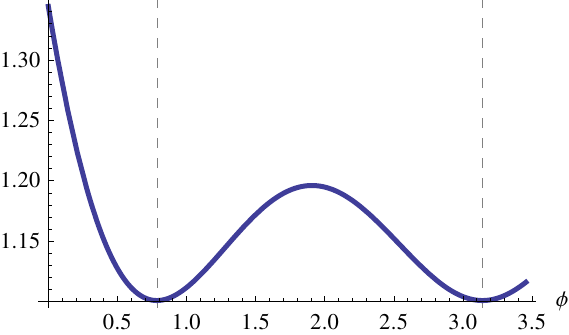}

(b)\includegraphics[width=7cm]{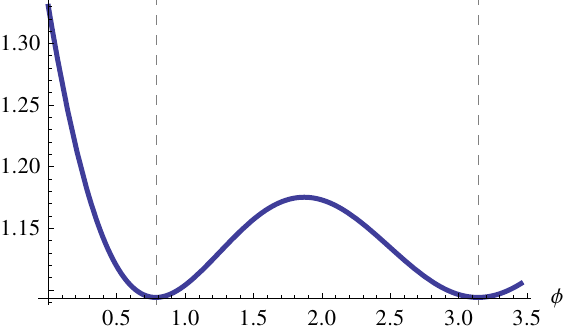}
\caption{The dependence of plaquette energy $E_{\rm plaq}(\phi)$ on the value of gauge-invariant phase difference $\phi$ on one of its bonds (a) for the case of the generalzied interaction $U_{\rm gen}(\phi)$ with $p=p_c$; (b) for the case of two-harmonics interaction $U_{\rm SFS}(\phi)$ with $t=t_c$.
}
\label{fig:Eplaq}
\end{figure}

Under the condition that the value of gauge-invariant phase difference $\phi$ on one of the bonds differs from that on three other bonds (for which the values of $\phi$ are assumed to be equal) the energy of a single square plaquette 
is given by
\begin{equation}                                              \label{Eplaq}
    E_{\rm plaq}(\phi)=U(\phi)+3U[(\pi-\phi)/3]\,.
\end{equation}
For $p<p_c$ function $E_{\rm plaq}(\phi)$ is minimal at $\phi=\pi/4$ (when $\phi=\pi/4$ on all four bonds), whereas for $p>p_c$ it is minimal at $\phi=\pi$
(when on three other bonds $\phi=0$). At $p=p_c$, the two minima have equal depths, see Fig. \ref{fig:Eplaq}(a). At this point the first-order phase transition between the two families of the ground states takes place.
Accordingly, the twofold discrete degeneracy of the ground states (existing at $p<p_c$), at $p>p_c$ is replaced by a much more developed discrete degeneracy which leads to a change in the classification of topological excitations (and in the nature of low temperature phase) and to the emergence of a different scenario for the disordering of the system.

\section{FF XY model with two-harmonics interaction}

Let us now check how the form of the interaction function evolves if instead of smoothly changing the exponent in Eq. (\ref{Ugen}) one adds to $U_0(\phi)$ the second harmonics,
\begin{equation}                                          \label{USFS-0}
    U_{\rm}(\phi)=
    J_1(1-\cos\phi)+J_2(1-\cos 2\phi)\,.
\end{equation}
The two-harmonics function (\ref{USFS-0}) can be used for the description of SFS junctions in the vicinity of so-called $0-\pi$ transition, at which the amplitude of the first harmonics 
changes sign \cite{Chtch01r,Chtch01,Golubov02r,Golubov02,Barash02}.
In this region of parameters, the amplitude of the main contribution to the interaction function [having the form $U_0(\phi)$] is strongly suppressed and may be of the same order as that of the next (second) harmonics.
Investigation of the conventional (without frustration) XY models with two-harmonics interaction of the form (\ref{USFS-0}) started long ago \cite{Korshunov85r,*Korshunov85,Lee85}
and became rather active in recent years, see Refs. \onlinecite{Shi,Hubscher13,Qi13} and references therein.

For further analysis it is convenient to rewrite Eq. (\ref{USFS-0}) with $J_{1,2}>0$ as
\begin{equation}                                          \label{USFS}
    U_{\rm SFS}(\phi)=
    \frac{J}{1+4t}\left[(1-\cos\phi)+t(1-\cos 2\phi) \right]\,,
\end{equation}
where $t=J_2/J_1$ denotes the ratio of the two amplitudes, and $J=J_1+4J_2$ is the effective coupling constant which is used in the figures as the unit of energy. Like in the case of $U_{\rm gen}(\phi)$, we have included into the definition of $U_{\rm SFS}(\phi)$ a dimensionless prefactor whose value is chosen to make $U_{\rm SFS}''(0)=J$ independent of parameter $t$ responsible for the {shape} of $U_{\rm SFS}(\phi)$.

\begin{figure}[b] \centering
\includegraphics[width=7cm]{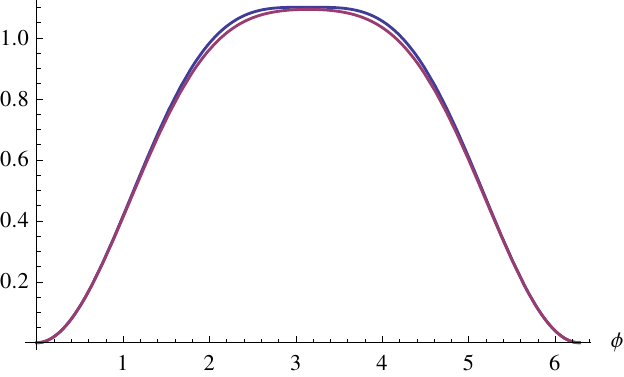}
\caption{(color online) Comparison of the form of $U_{\rm gen}(\phi)$ at $p=p_c$ (upper curve) and that of $U_{\rm SFS}(\phi)$ for $t=t_c$ (lower curve).
}
\label{fig:comparison3}
\end{figure}

Fig. \ref{fig:comparison}(b) demonstrates how the increase of the ratio of the amplitudes of the second and first harmonics  $t$ changes the form of function $U_{\rm SFS}(\phi)$. Comparison of Fig. \ref{fig:comparison}(b) with Fig. \ref{fig:comparison}(a) shows that the increase of parameter $t$ influences the form of the interaction function basically in the same way as the increase of parameter $p$ in Eq. (\ref{Ugen}). In particular, the value of $U_{\rm SFS}(\pi)$ is suppressed with the increase in $t$ much stronger than that of $U_{\rm SFS}(\pi/4)$. This suggests that  the FF XY model with interaction (\ref{USFS}) can be expected to demonstrate the reconstruction of the ground state structure completely analogous to the one in the generalized FF XY model of Ref. \onlinecite{Minnhagen07}. Indeed, this reconstruction takes place at
\[
    t_c\approx 0.207108\,,
\]
where the two minima of the function $E_{\rm plaq}(\phi)$ defined by Eq. (\ref{Eplaq}) have equal depths, see Fig. \ref{fig:Eplaq}(b). Like in the case of $U_{\rm gen}(\phi)$ with {$p>p_c$}, at $t>t_c$ the energy of a plaquette is minimal when on one of its bonds $\phi=\pi$ and on the other three bonds $\phi=0$.

To further emphasize the similarity of the two models, in Fig. \ref{fig:comparison3} we compare the form of $U_{\rm gen}(\phi)$ at $p=p_c$
and that of $U_{\rm SFS}(\phi)$ at $t=t_c$. This graph demonstrates that these two functions are very close to each other. This is not a mere coincidence.
Although for an arbitrary $p$ function $U_{\rm gen}(\phi)$ defined by Eq. (\ref{Ugen}) looks like an unphysical one, at
\begin{equation}                                                   \label{}
    p=p_*=\sqrt{2}\approx 1.414214
\end{equation}
it exactly coincides with function $U_{\rm SFS}(\phi)$ defined by Eq. (\ref{USFS}) with
\begin{equation}
 t=t_*=0.25\,.
\end{equation}
Since $p_c$ is relatively close to $p_*$, and $t_c$ to $t_*$,
\begin{equation}                                                    \label{}
p_*-p_c\approx 0.160(p_*-1)\,,~~~t_*-t_c\approx 0.172t_*\,,
\end{equation} it is no wonder that the two functions plotted in Fig. \ref{fig:comparison3} are very close to each other.

\begin{figure}[t] \centering
\includegraphics[width=7cm]{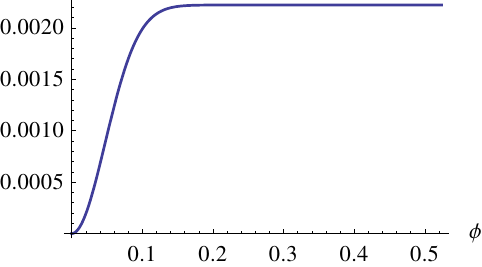}
\caption{The form of $U_{\rm gen}(\phi)$ at $p=30$. Note an extended scale on the horizontal axis.
}
\label{fig:Ugen30}
\end{figure}
At this point it has to be emphasized that the most important features  of the phase diagram of the generalized FF XY model  (namely, the line separating the two low-temperature phases from each other and three multicritical points) are restricted to the narrow ($<1\%$) interval around $p_c$, see Fig. 5 of Ref. \onlinecite{Minnhagen08}.  The extreme closeness of the two interaction functions depicted in Fig. \ref{fig:comparison3} allows one to expect the existence of analogous features in the phase diagram of the magnetically frustrated SFS array in a narrow vicinity of the line
\begin{equation}                                                \label{}
    \frac{J_1(T)}{J_2(T)}=\frac{1}{t_c}\approx 4.828\,.
\end{equation}

Since the two models have the same classification of topological excitations in both of their low-temperature phases (below and above $p_c$ or $t_c$), the general topology of their phase diagrams can also be expected to be the same. However, one of the features of the phase diagram of the generalized FF XY model, namely, the transformation of the phase transition from a continuous to a first-order one taking place at $p\approx 30$ \cite{Minnhagen08}, cannot be reproduced in the model with two-harmonics interaction (\ref{USFS}) because the form of $U_{\rm gen}(\phi)$ at $p\approx 30$ is extremely flat outside of a very narrow vicinity of $\phi=0$ (see Fig. \ref{fig:Ugen30}) and in no way can be approximated by the two-harmonics function $U_{\rm SFS}(\phi)$.

\begin{figure}[t] \centering
\includegraphics[width=7cm]{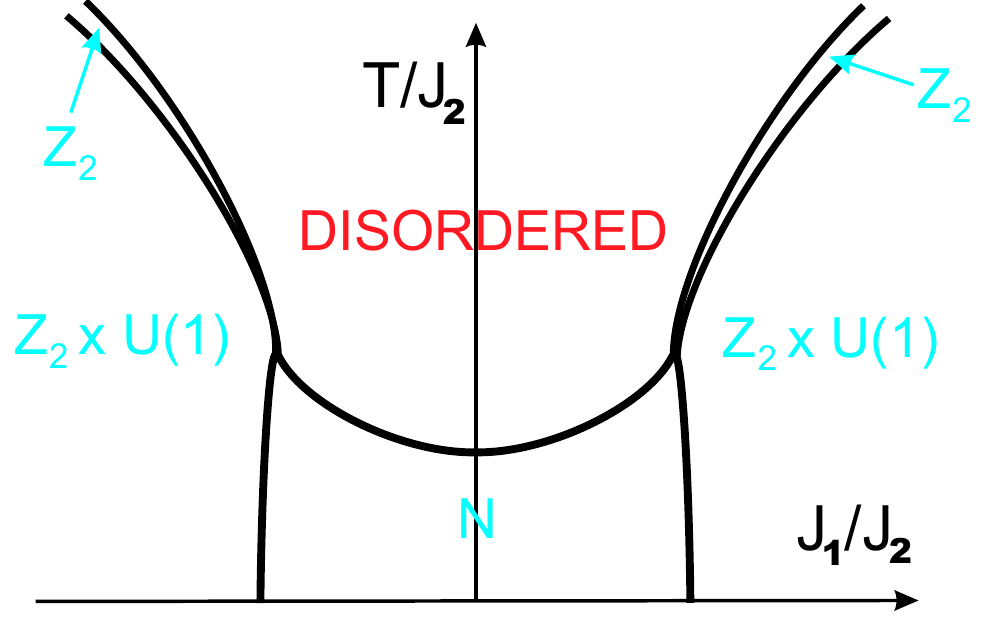}
\caption{(color online) The schematic structure of the phase diagram of a fully frustrated array of SFS junctions with two-harmonics interaction (\ref{USFS-0}). The phase denoted Z$_2\times$U(1) combines a spontaneously broken Z$_2$ symmetry with a finite value of helicity modulus, whereas in the phase denoted Z$_2$ the helicity modulus is equal to zero. The phase with a modified structure of the ground states (leading to their extensive degeneracy) is denoted N.
}
\label{fig:ph-diagram}
\end{figure}

\vspace*{-2mm}
\section{Conclusion}

To conclude, in the present note we have shown that a magnetically frustrated square array of SFS junctions \cite{Ryazanov01,Golubov04,Frolov08} with half-integer number of flux quanta per plaquette has to demonstrate a rich phase diagram with the structure quite analogous to that of the generalized FF XY model \cite{Minnhagen07,Minnhagen08} with interaction (\ref{Ugen}).
Like in the case of the unfrustrated SFS array with square lattice \cite{Korshunov10}, the phase diagram of the fully frustrated square SFS array has to be symmetric with respect to the change of the sign of parameter $J_1$ in Eq. (\ref{USFS-0}) at constant value of $J_2$. This follows from the possibility to shift in Hamiltonian (\ref{H}) all variables $A_{jj'}$ by $\pi$ without violating the constraint (\ref{f}). Accordingly, in the phase diagram of the FF square SFS array the phase diagram of the generalized FF XY model has to be reproduced {\em twice}, on both sides of the $0-\pi$ transition.

The schematic structure of this phase diagram is shown in Fig. \ref{fig:ph-diagram}, where it is plotted in variables $J_1/J_2$ and $T/J_2$. In addition to the three phases which exist in the conventional fully frustrated XY model \cite{Korshunov06} (corresponding to $J_2\rightarrow 0$ limit) this phase diagram includes also an additional phase denoted $N$. Although in Ref. \onlinecite{Minnhagen08} the properties of this phase are characterized with the help of a rather complicated order parameter (involving counting the number of corners on domain walls) we expect that it can be more transparently described by comparing the behavior of correlation functions for variables $\exp i\theta$ and $\exp 2i\theta$, which is relegated to a separate publication.

A frustrated XY model with square lattice and two-harmonics interaction has been studied earlier in Ref. \onlinecite{Qin09}. However, for some unexplained reason, the authors of this work have chosen to consider an interaction which has not gauge-invariant form, which raises doubts on applicability of their results for describing physical systems. The model suitable for the description of FF SFS array with triangular lattice has been studied in Ref. \onlinecite{Park08}, however this work considers the case which in the notation of our Eq. (\ref{USFS-0}) corresponds to $J_2<0$ and cannot be compared to the case $J_2>0$ considered here.

\vspace*{-2mm}
\acknowledgments

This work has been supported by Russian Science Foundation through grant No. 14-12-00898.

\bibliography{bibliography-sfs-ff}

\end{document}